\documentclass[epjCONF]{svjour}
\usepackage{graphics}
\ifx\pdfoutput\undefined
  \usepackage{graphicx}
\else
  \usepackage[pdftex]{graphicx}
  \usepackage{epstopdf}
\fi
\usepackage[varg]{txfonts} 
\usepackage[latin1]{inputenc}
\session-title{MESON2012}
\begin{document}
%
\title{Observation of a $J^{PC}=1^{-+}$ exotic signal in the $\pi^{-}\pi^{0}\pi^{0}$ system diffractively produced at COMPASS, 
and comparison to the charged decay mode
}
\author{Frank Nerling\thanks{\email{nerling@cern.ch}}, for the COMPASS collaboration}
\institute{Physikalisches Institut, Albert-Ludwigs-Universit\"at Freiburg / CERN PH Department, Geneva}
\abstract{
The COMPASS experiment at the CERN SPS features good charged particle tracking and coverage by electromagnetic 
calorimetry, and our data provide excellent opportunity for simultaneous observation of new states in two different 
decay modes within the same experiment. The existence of the spin-exotic $\pi_1(1600)$ resonance in the $\rho\pi$ 
decay channel is studied for the first time in COMPASS in both decay modes of the diffractively produced $(3\pi)^{-}$ 
system: $\pi^{-}p \rightarrow \pi^{-}\pi^{0}\pi^{0}p$ ~and~ $\pi^{-}p \rightarrow \pi^{-}\pi^{+}\pi^{-}p$. 
A preliminary partial-wave analysis (PWA) performed on the 2008 proton target data allows for a first conclusive comparison 
of both $(3\pi)^{-}$ decay modes not only for main waves but also for small ones, including the spin-exotic $1^{-+}$ 
wave. 
We find the neutral versus charged mode results in good agreement with expectations from isospin symmetry. Both, the 
intensities and the relative phases to well-known resonances, are consistent for the neutral and the charged decay 
modes of the $(3\pi)^{-}$ system. The status on the search for the spin-exotic $\pi_1(1600)$ resonance produced on a 
proton target is discussed.
} 
\maketitle
\section{Introduction}
\label{intro}
\vspace{-0.3cm}
The $\pi_1(1600)$ is a hybrid meson candidate that has been reported by different experiments and in different 
decay channels. The experimental observation of such a resonance beyond the simple Constituent Quark Model
would be a fundamental confirmation of Quantum Chromodynamics, allowing for and predicting such spin-exotic 
mesons according to various models, for a recent overview, see e.g.~\cite{MeyerHaarlem2010}.
The observations in the $\rho\pi$ decay channel analysed in 3$\pi$ final states are still controversially 
discussed in the community. Especially the resonant nature of the observed signals in the exotic $J^{PC}=1^{-+}$ 
partial-wave reported by the E852 at BNL and the VES experiments~\cite{Adams:1998,Khokhlov:2000} in diffractively produced 
$\pi^{-}\pi^{+}\pi^{-}$ final states are questioned, in later publications previous conclusions were withdrawn~\cite{Amelin:2005} 
and re-analyses of the $(3\pi)^{-}$ system in the charged and neutral decay modes led to opposite conclusions~\cite{Dzierba:2006}. 
One may get a hint at this controversy looking at~\cite{PDG}. 

In the 2004 pilot run data, COMPASS observed a significant $J^{PC}$ spin-exotic signal in diffractively produced 
three charged pion final states on a Pb target at $1660$$\pm$$10^{+0}_{-64}$\,MeV/c$^2$ that is consistent with the 
disputed $\pi_1(1600)$; it shows a clean phase motion against well-known resonances~\cite{Alekseev:2009a}. 
The present results from the high statistics 2008 proton target data discussed here were obtained employing the same 
PWA model as in~\cite{Alekseev:2009a}. The results are consistent with the previous observations. Apart of the prominent 
and established resonances $a_1(1260)$, $a_2(1320)$, $\pi_2(1670)$, and $\pi(1800)$, $a_4(2040)$, we observe an exotic 
signal in the $1^{-+}$ wave at around 1.6\,GeV/$c^2$ that shows a clean, rapid phase motion with respect to well-known 
resonances. Not only the intensities of the observed resonances and the exotic wave but also the relative phase differences 
are observed consistently for both, the neutral and the charged decay modes of the $\rho\pi$ decay channel.

\begin{figure}[tp!]
    \begin{center}
      \vspace{-0.5cm}
      \includegraphics[clip,trim= 110 120 60 220,width=1.0\linewidth]
      {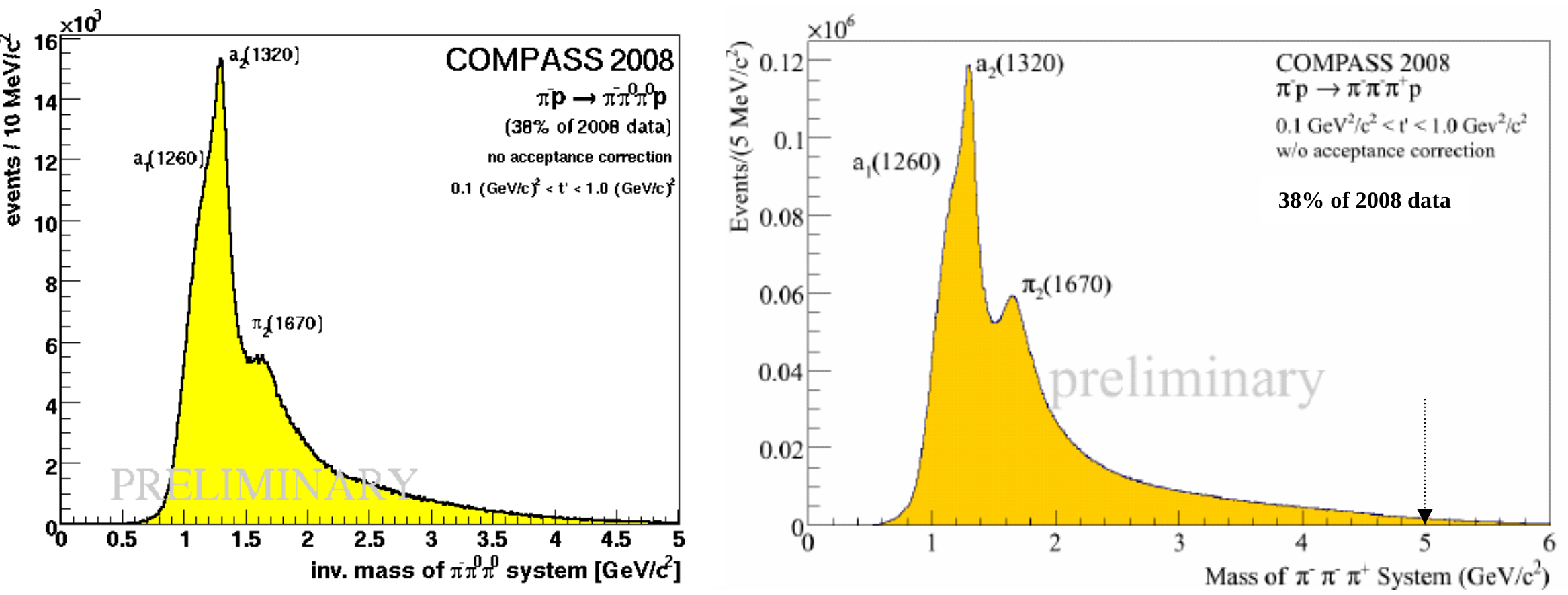}
      \vspace{-0.5cm}
      \caption{Total mass spectrum of the $(3\pi)^{-}$ systems --- neutral \textit{(left)} versus charged decay mode \textit{(right)}.
The spectra look similar as expected, showing both the most prominent, well-known resonances $a_1(1260)$, $a_2(1320)$ and $\pi_2(1670)$.}
      \label{fig:3piMassTot_neutral_charged}
      \vspace{-0.4cm}
    \end{center}
\end{figure}

\section{Partial-wave analysis results}
\label{sec:1}
\vspace{-0.3cm}
%
%
\begin{figure}[bp!]
  \begin{minipage}[h]{.32\textwidth}
    \begin{center}
         \vspace{-0.5cm}
\resizebox{1.0\columnwidth}{!}{%
  \includegraphics[clip,trim= 5 0 10 15, width=1.0\linewidth, angle=0]{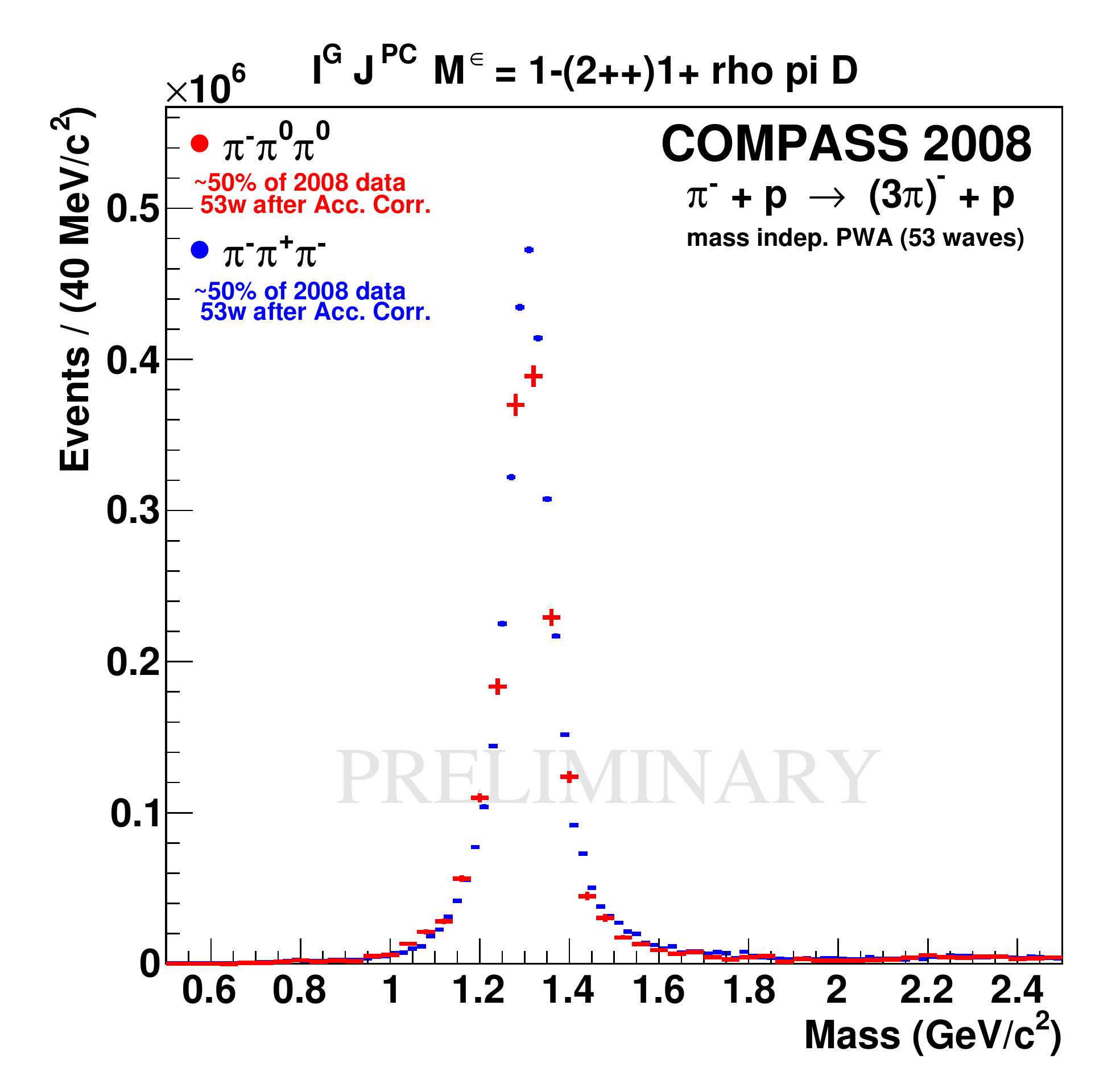} }
    \end{center}
  \end{minipage}
  \hfill
  \begin{minipage}[h]{.32\textwidth}
    \begin{center}
      \vspace{-0.5cm}
\resizebox{1.0\columnwidth}{!}{%
     \includegraphics[clip,trim= 5 0 10 15, width=1.0\linewidth, angle=0]{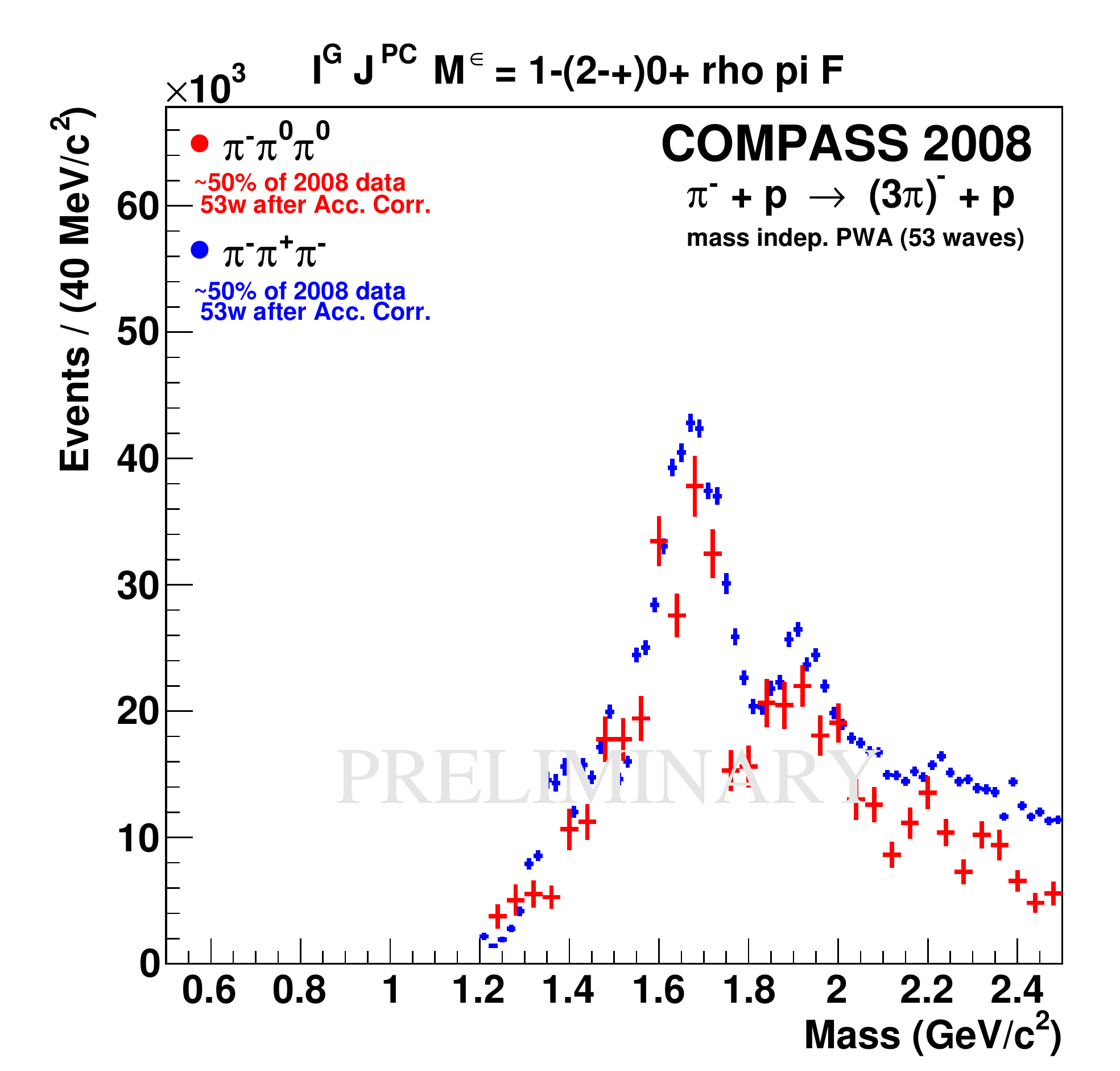} }
    \end{center}
  \end{minipage}
  \begin{minipage}[h]{.32\textwidth}
    \begin{center}
      \vspace{-0.5cm}
\resizebox{1.0\columnwidth}{!}{%
     \includegraphics[clip,trim= 5 -5 10 20, width=1.0\linewidth, angle=0]{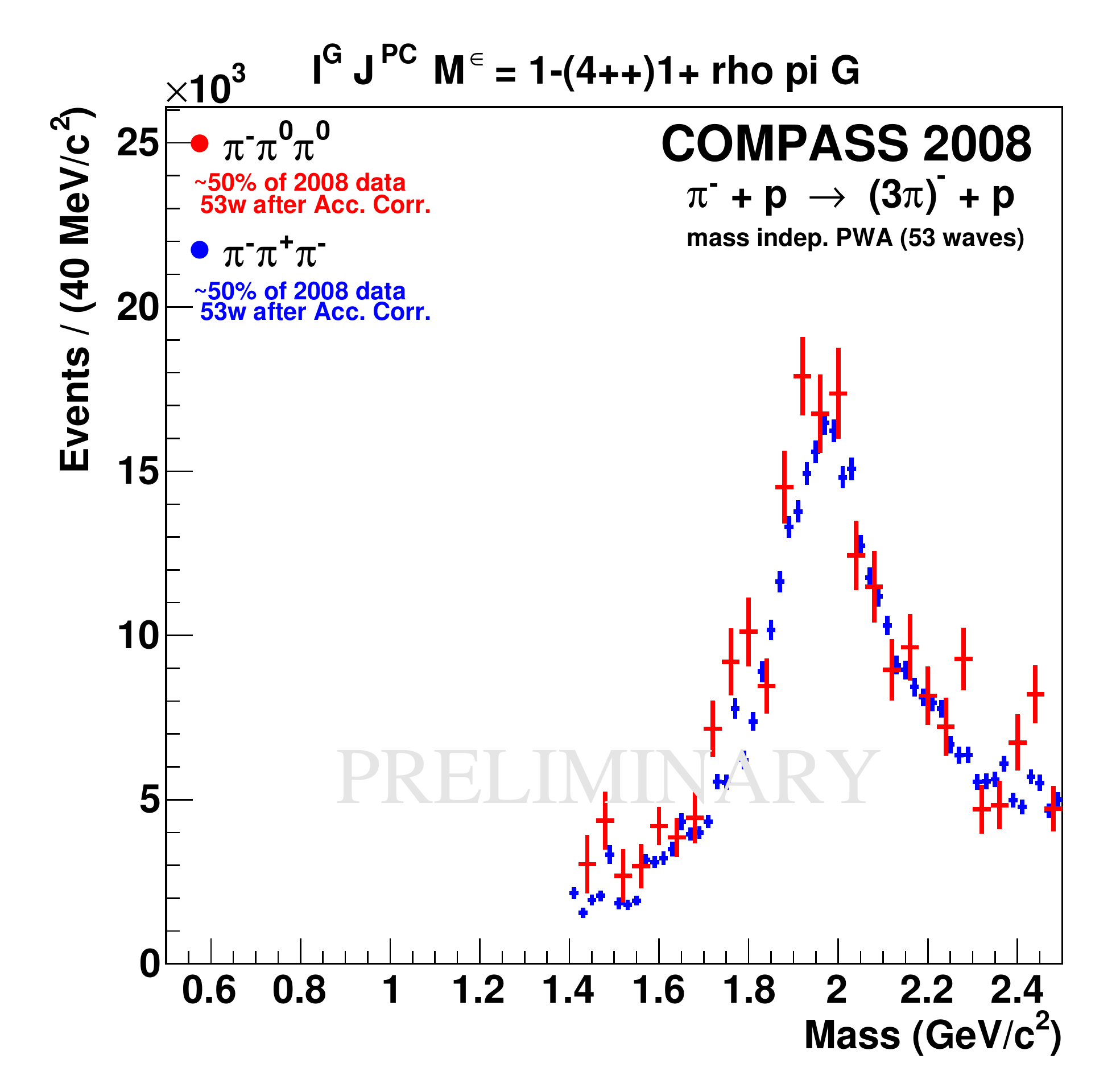}}
    \end{center}
  \end{minipage}
  \begin{minipage}[h]{.32\textwidth}
    \begin{center}
      \vspace{-0.52cm}
\resizebox{1.0\columnwidth}{!}{%
  \includegraphics[clip,trim= 5 0 10 15, width=1.0\linewidth, angle=0]{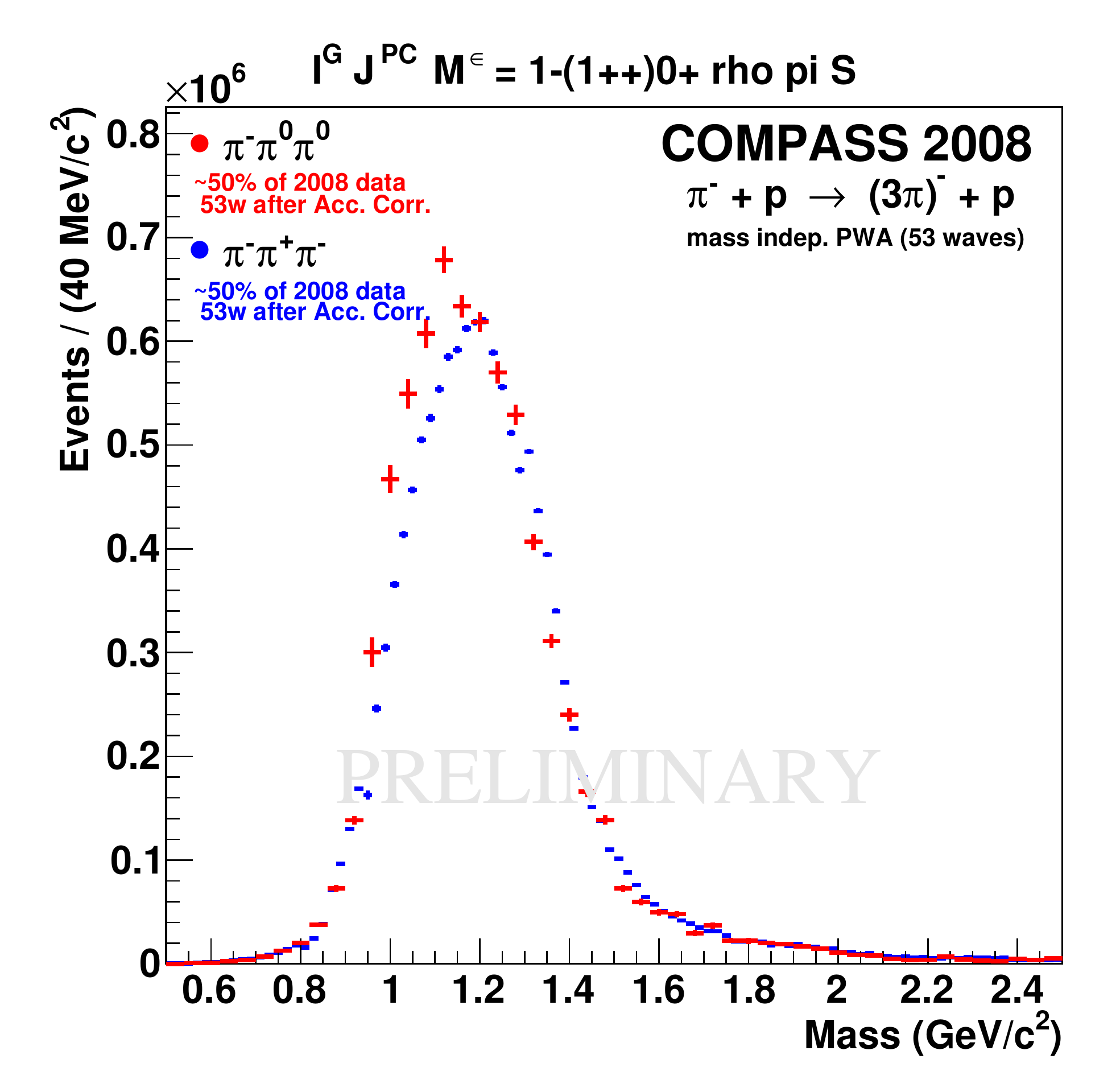} }
    \end{center}
  \end{minipage}
  \hfill
  \begin{minipage}[h]{.32\textwidth}
    \begin{center}
      \vspace{-0.52cm}
\resizebox{1.0\columnwidth}{!}{%
     \includegraphics[clip,trim= 5 0 10 15, width=1.0\linewidth, angle=0]{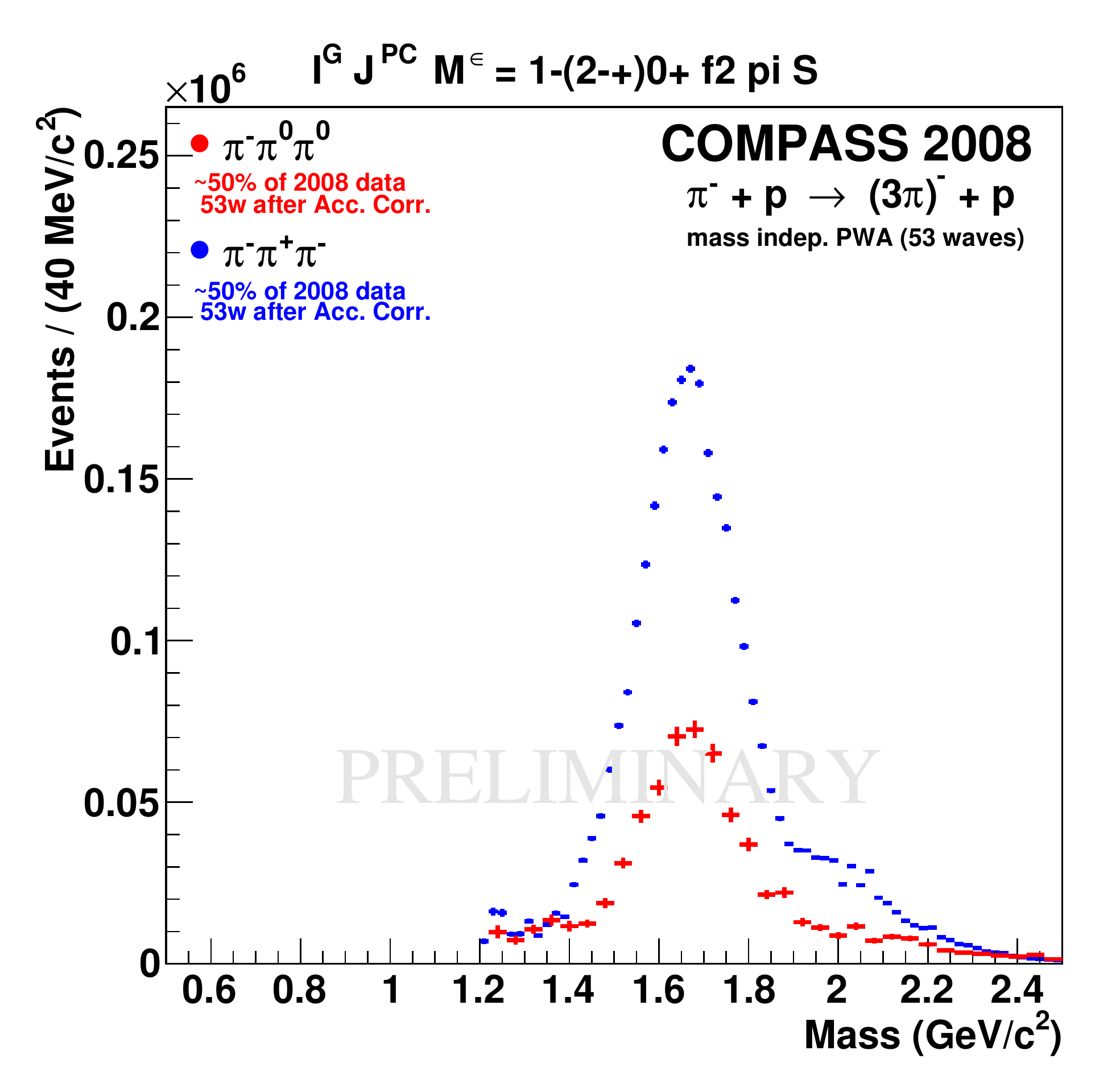} }
    \end{center}
  \end{minipage}
  \begin{minipage}[h]{.32\textwidth}
    \begin{center}
      \vspace{-0.5cm}
\resizebox{1.0\columnwidth}{!}{%
     \includegraphics[clip,trim= 5 0 10 15, width=1.0\linewidth, angle=0]{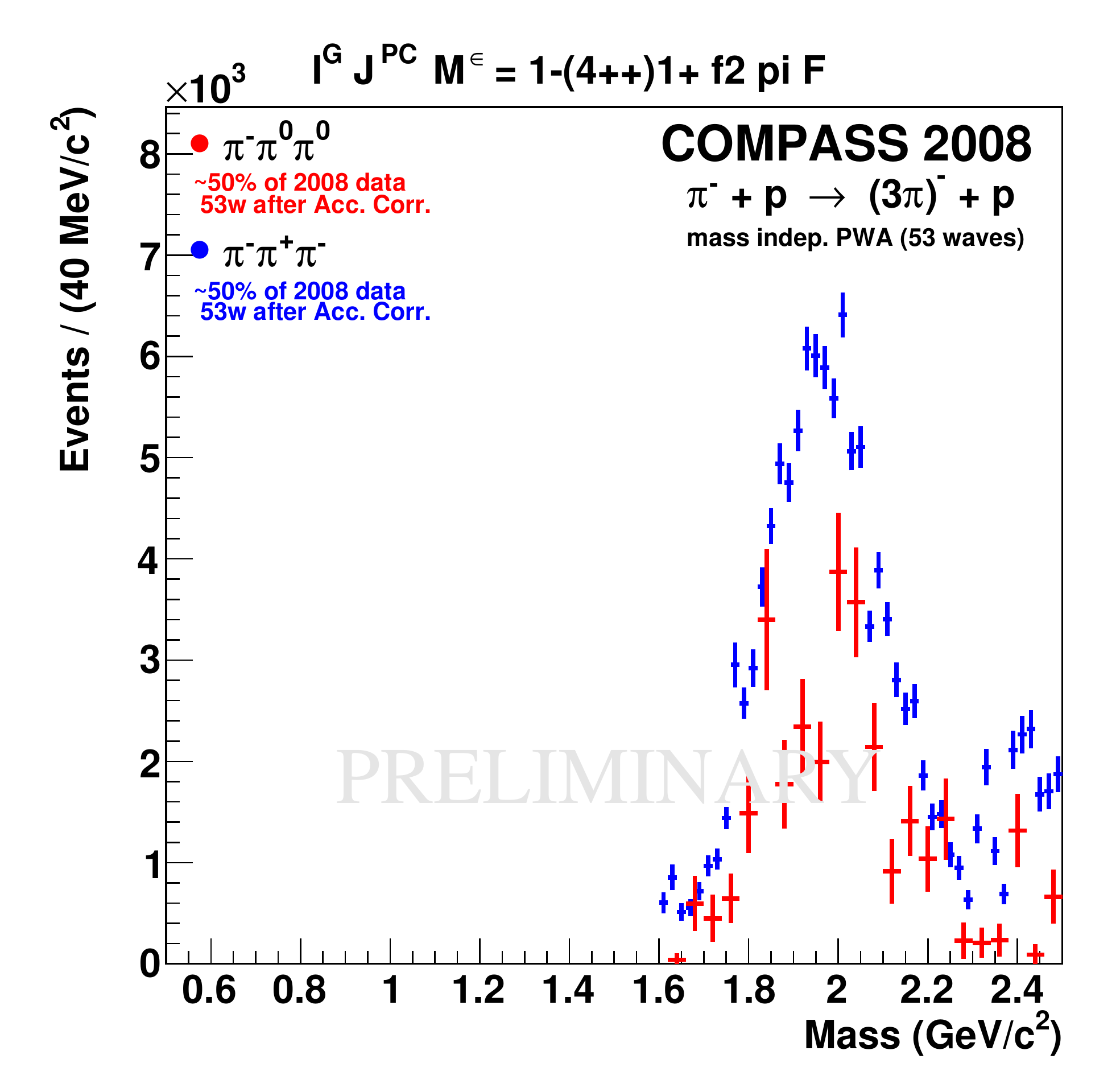}}
    \end{center}
  \end{minipage}
    \begin{center}
      \vspace{-0.3cm}
     \caption{Mass-independent PWA result for neutral (red) versus charged (blue) mode 
       --- major and small waves. The prominent and established resonances are consistently observed: 
       the $a_2(1320)$ {\it (top, left)} and the $a_1(1260)$ {\it (bottom, left)}, and 
       the $\pi_2(1670)$ {\it (top \& bottom, centre)} and the $a_4(2040)$ {\it (top \& bottom, right)} decaying 
       into $\rho\pi$ and $f_2\pi$.
}
       \label{fig:isospinSymmMainWaves}
     \end{center}
     \vspace{-0.7cm}
\end{figure}
\begin{figure}[tp!]
  \begin{minipage}[h]{.32\textwidth}
    \begin{center}
      \vspace{-0.5cm}
\resizebox{1.0\columnwidth}{!}{%
  \includegraphics[clip,trim= 5 0 10 15, width=1.0\linewidth, angle=0]{Plots/PWA_Neutral_vs_Charged/h32.pdf} }
    \end{center}
  \end{minipage}
  \hfill
  \begin{minipage}[h]{.32\textwidth}
    \begin{center}
      \vspace{-0.5cm}
\resizebox{1.0\columnwidth}{!}{%
     \includegraphics[clip,trim= 5 0 10 15, width=1.0\linewidth, angle=0]{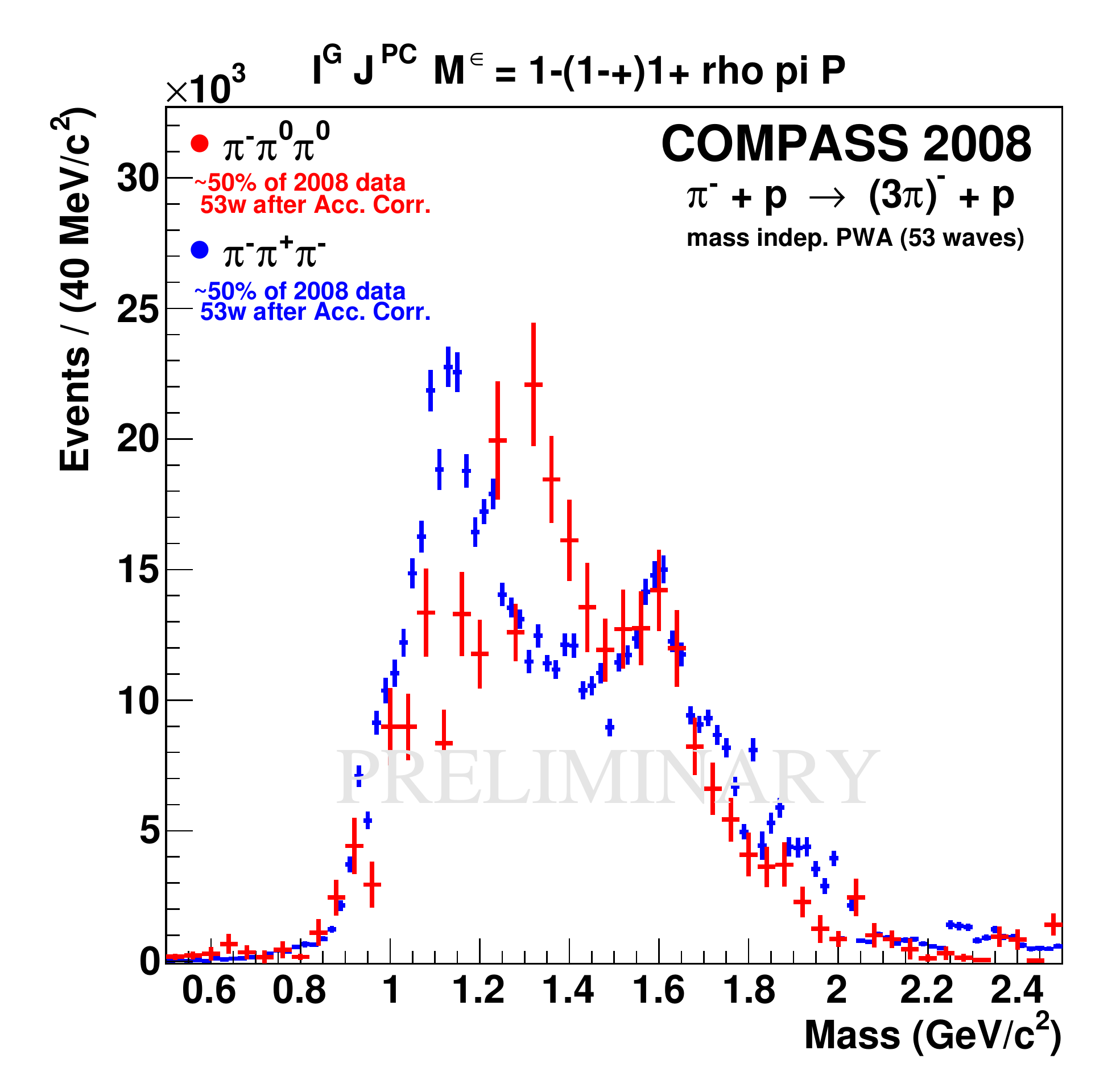} }
    \end{center}
  \end{minipage}
  \begin{minipage}[h]{.32\textwidth}
    \begin{center}
      \vspace{-0.5cm}
\resizebox{1.0\columnwidth}{!}{%
     \includegraphics[clip,trim= 5 -5 10 20, width=1.0\linewidth, angle=0]{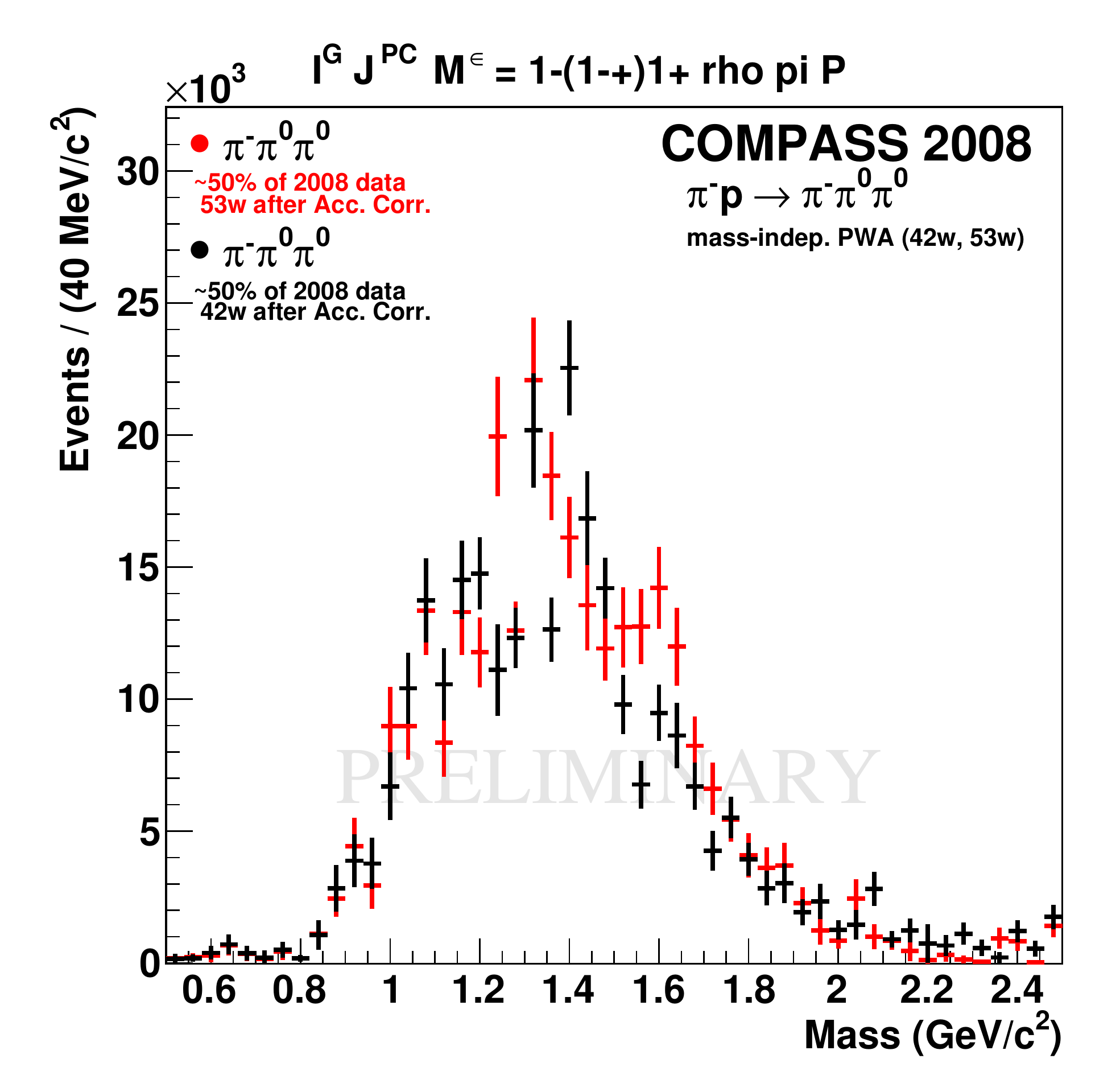}}
    \end{center}
  \end{minipage}
  \begin{minipage}[h]{.32\textwidth}
    \begin{center}
      \vspace{-0.51cm}
\resizebox{1.0\columnwidth}{!}{%
  \includegraphics{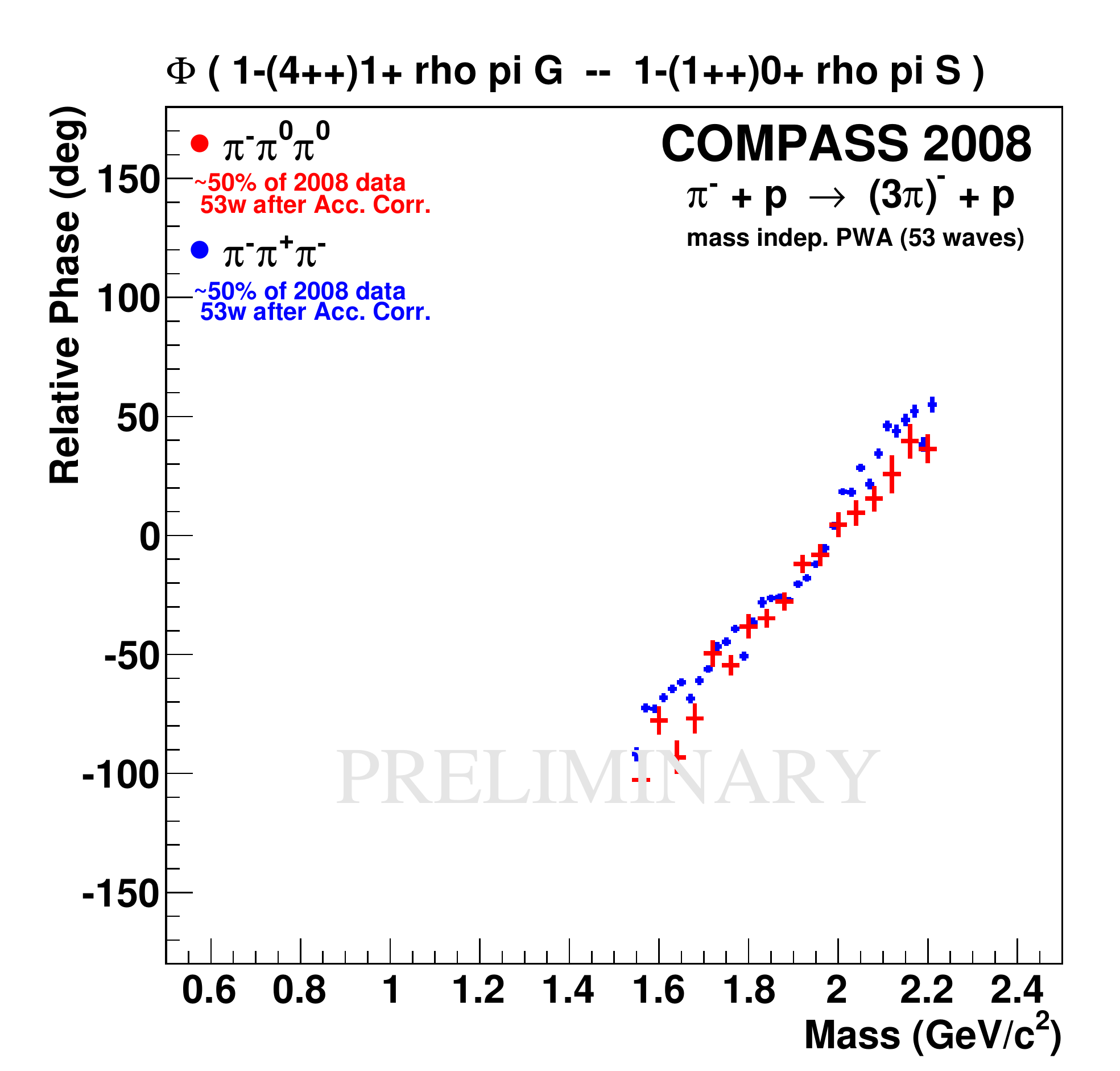} }
    \end{center}
  \end{minipage}
  \hfill
  \begin{minipage}[h]{.32\textwidth}
    \begin{center}
      \vspace{-0.52cm}
\resizebox{1.0\columnwidth}{!}{%
     \includegraphics[clip,trim= 5 0 10 15, width=1.0\linewidth, angle=0]{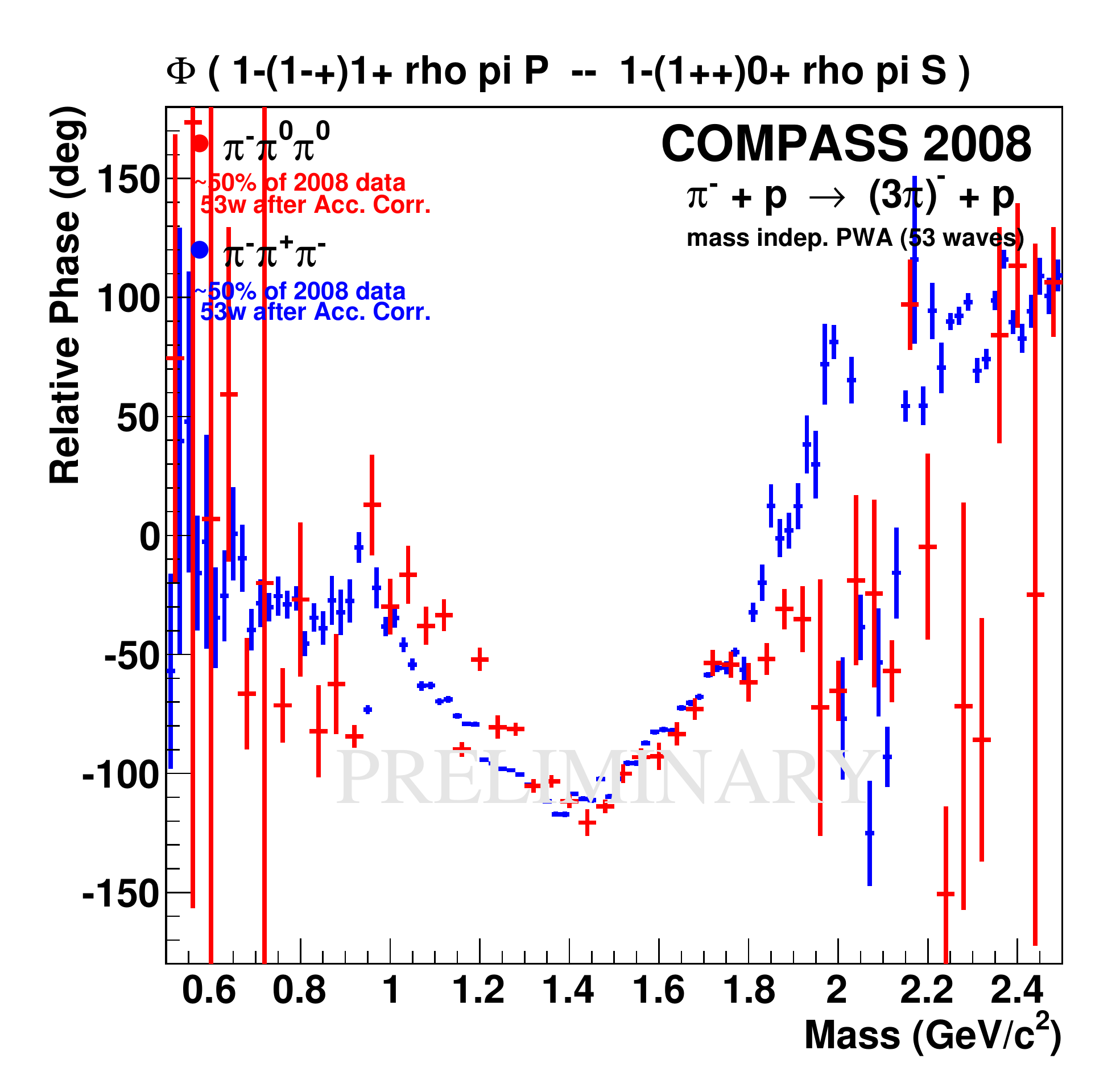} }
    \end{center}
  \end{minipage}
  \begin{minipage}[h]{.32\textwidth}
    \begin{center}
      \vspace{-0.5cm}
\resizebox{1.0\columnwidth}{!}{%
     \includegraphics[clip,trim= 5 0 10 15, width=1.0\linewidth, angle=0]{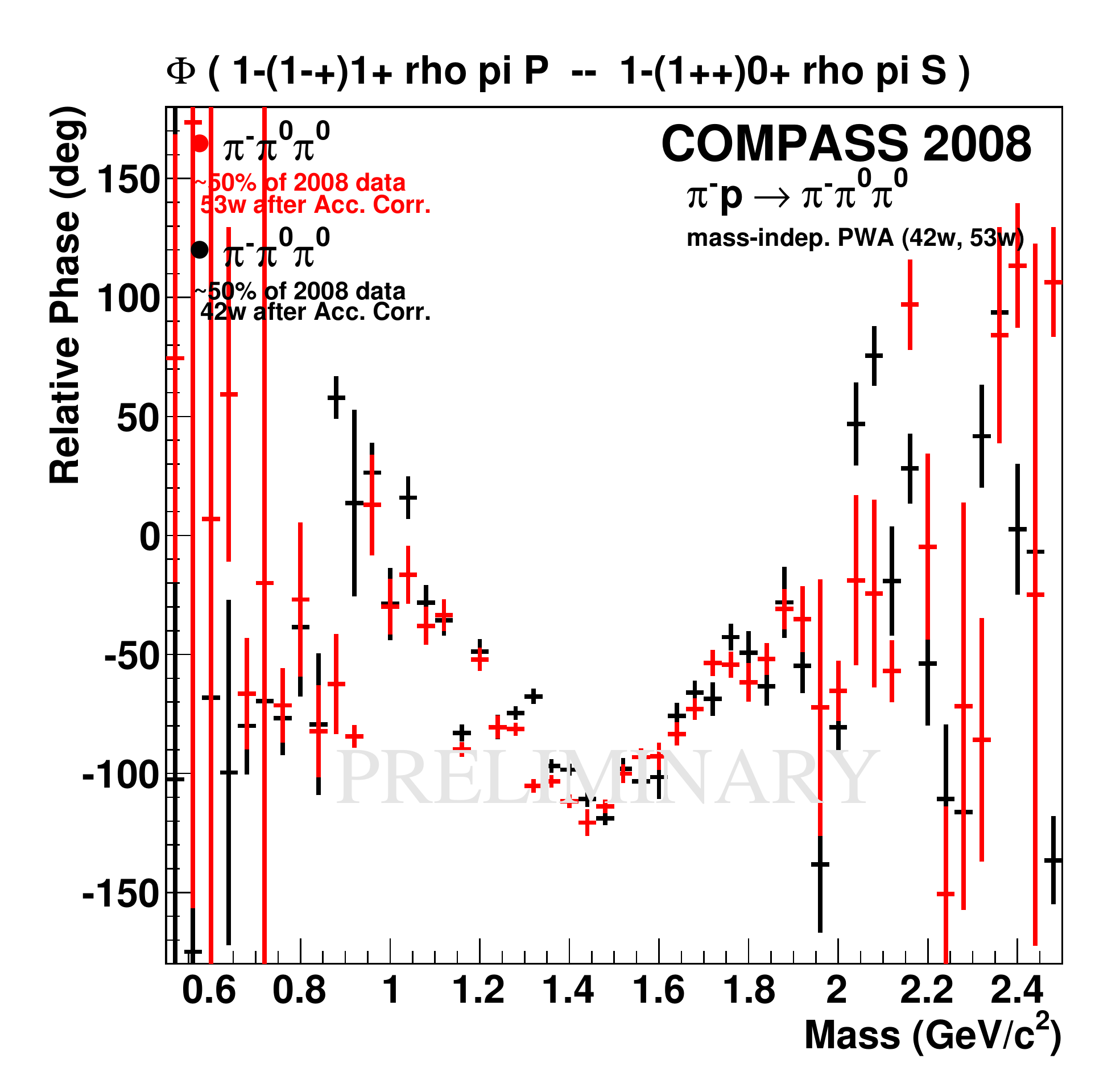}}
    \end{center}
  \end{minipage}
    \begin{center}
     \vspace{-0.3cm}
     \caption{Mass-independent PWA result for the exotic $1^{-+}$ wave.
       The $(4^{++})1^+\,\rho\pi\,G$ wave (of comparable intensity) showing the $a_4(2040)$ is displayed again 
       for comparison {\it (top, left)}, the $(1^{-+})1^+\,\rho\pi\,P$ wave is shown for the neutral mode data (red)
       compared to the charged one (blue) {\it (top, centre)}, and also for two different wave-sets {\it (top, right)}: 
       the initial one (black) comprising 42 partial-waves~\cite{Alekseev:2009a,nerling:2009} and the one extended by 11 
       additional waves. The phase differences with respect to the $(1^{++})0^+ \rho\pi S$ wave are given below, 
       respectively.}  
        \label{fig:exotic}
     \end{center}
     \vspace{-0.9cm}
\end{figure}

A PWA using a similar model as employed in~\cite{Alekseev:2009a} has been performed on about 
50\,\% of the 2008 190\,GeV/$c$ $\pi^{-}$ beam data for the neutral and charged decay mode data 
(Fig.\,\ref{fig:3piMassTot_neutral_charged}). Details on the analyses are given in~\cite{nerling:2011} and~\cite{haas:2011}, 
respectively, the PWA method applied is summarised in~\cite{nerling:2009}.  
The mass-independent PWA results using a wave-set of 53 partial-waves (the 42 wave-set used previously~\cite{Alekseev:2009a} 
extended by 11 additional waves to account for the higher statistics analysed) are shown, for both decay modes after acceptance corrections applied in Figs.\,\ref{fig:isospinSymmMainWaves} -\,\ref{fig:exotic} for various selected waves.  

The most prominent resonances (Fig.\,\ref{fig:isospinSymmMainWaves}) are consistently observed in both decay modes in 
the major waves given in Fig.\,\ref{fig:isospinSymmMainWaves}. The $a_2(1320)$ and the $a_1(1260)$ decaying into $\rho\pi$ 
are observed with same width and intensity for both modes, similarly for the $\pi_1(1670)$ and $a_4(2040)$ decays into 
$\rho\pi$ (Fig.\,\ref{fig:isospinSymmMainWaves}, top/centre and top/right), whereas a suppression factor of about two is 
observed for the neutral mode intensities as compared to the charged mode data for the resonances decaying into $f_2\pi$ 
(Fig.\,\ref{fig:isospinSymmMainWaves}, bottom/centre and bottom/right) --- as expected. The data are found in good agreement 
with expectations from isospin symmetry (different yields for neutral vs. charged mode as expected from the involved 
Clebsch-Gordon coefficients and also taking into account Bose-Symmetrisation, see detailed discussion in~\cite{nerling:2011}), 
throughout the full wave-set, for partial-waves of large, small and very small intensities.  

\begin{figure}[tp!]
  \begin{minipage}[h]{.32\textwidth}
    \begin{center}
      \vspace{-0.5cm}
\resizebox{1.0\columnwidth}{!}{%
  \includegraphics[clip,trim= 5 0 10 15, width=1.0\linewidth, angle=0]{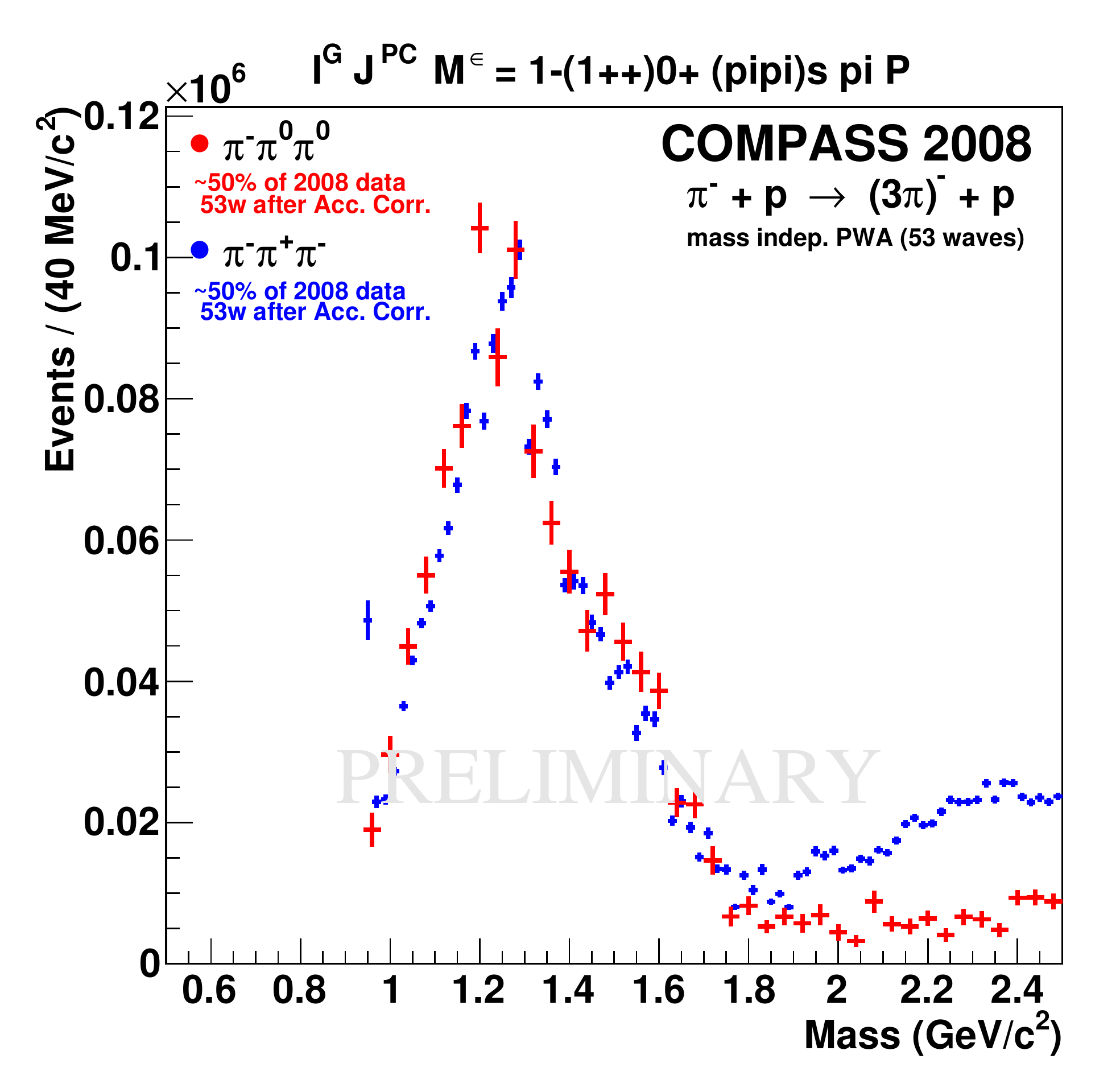} }
    \end{center}
  \end{minipage}
  \hfill
  \begin{minipage}[h]{.32\textwidth}
    \begin{center}
      \vspace{-0.5cm}
\resizebox{1.0\columnwidth}{!}{%
     \includegraphics[clip,trim= 5 0 10 15, width=1.0\linewidth, angle=0]{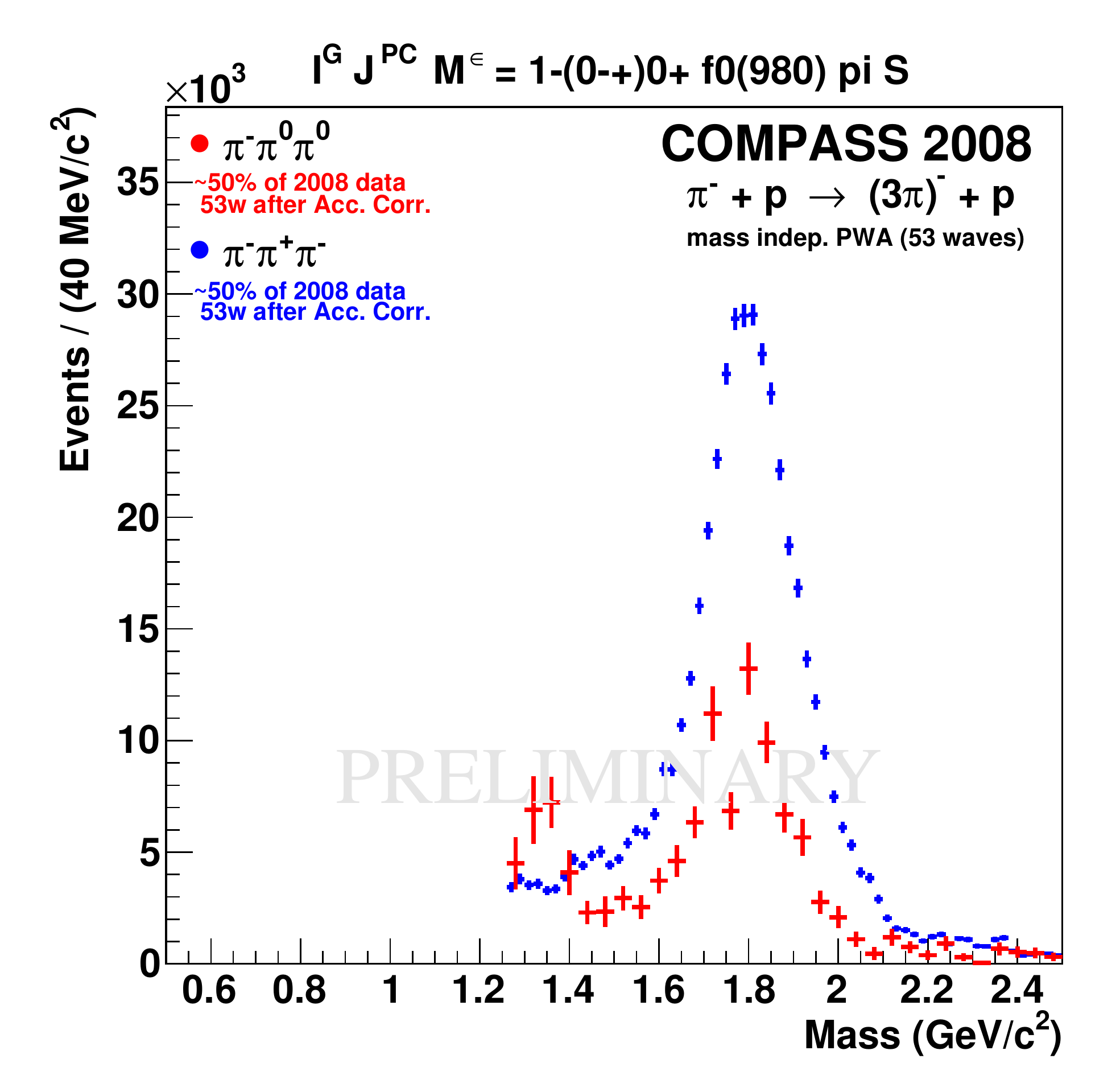} }
    \end{center}
  \end{minipage}
  \begin{minipage}[h]{.32\textwidth}
    \begin{center}
      \vspace{-0.5cm}
\resizebox{1.0\columnwidth}{!}{%
     \includegraphics[clip,trim= 5 -5 10 20, width=1.0\linewidth, angle=0]{Plots/PWA_Neutral_vs_Charged/h31.pdf}}
    \end{center}
  \end{minipage}
  \begin{minipage}[h]{.32\textwidth}
    \begin{center}
      \vspace{-0.52cm}
\resizebox{1.0\columnwidth}{!}{%
  \includegraphics[clip,trim= 5 0 10 15, width=1.0\linewidth, angle=0]{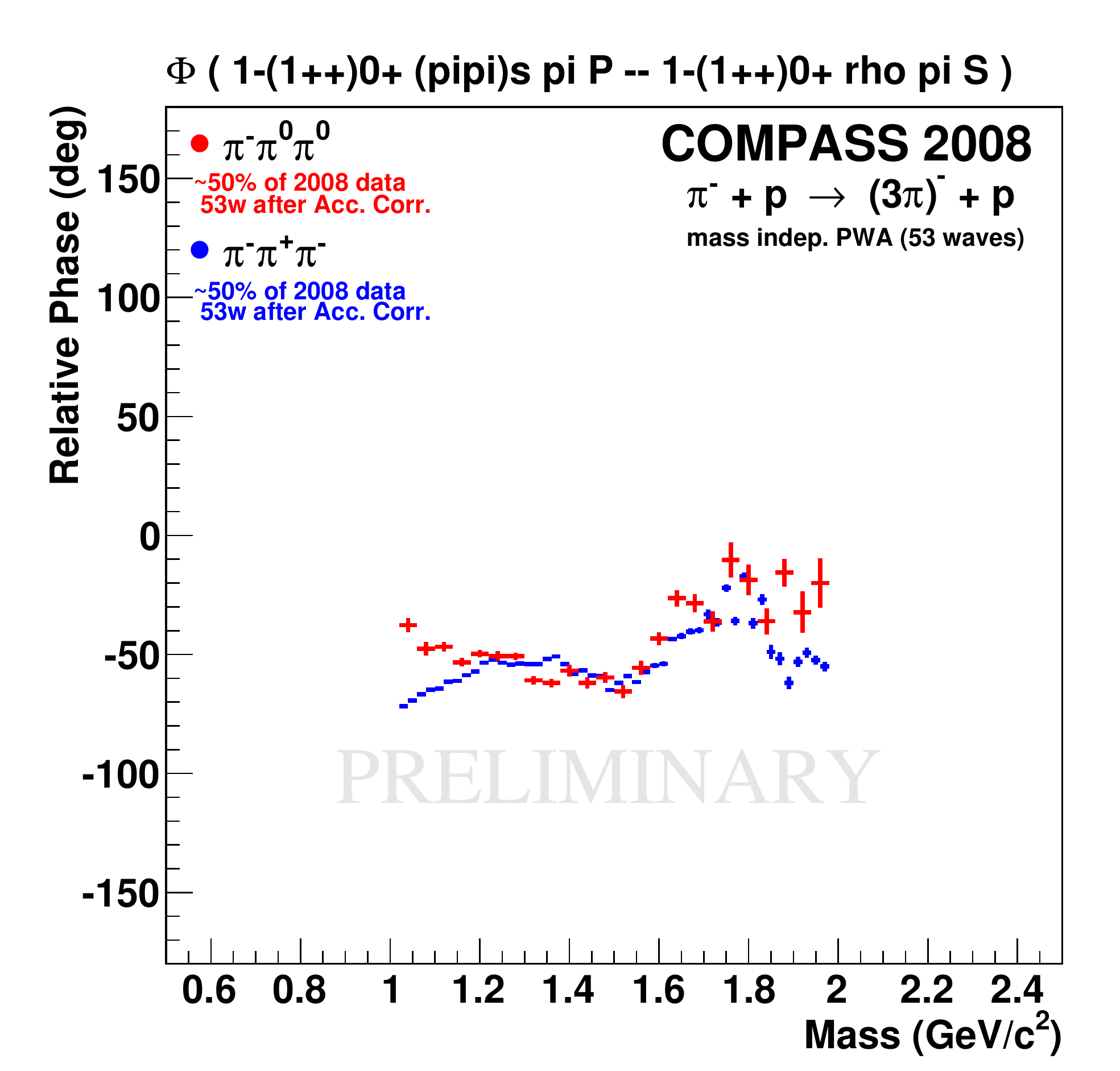} }
    \end{center}
  \end{minipage}
  \hfill
  \begin{minipage}[h]{.32\textwidth}
    \begin{center}
      \vspace{-0.52cm}
\resizebox{1.0\columnwidth}{!}{%
     \includegraphics[clip,trim= 5 0 10 15, width=1.0\linewidth, angle=0]{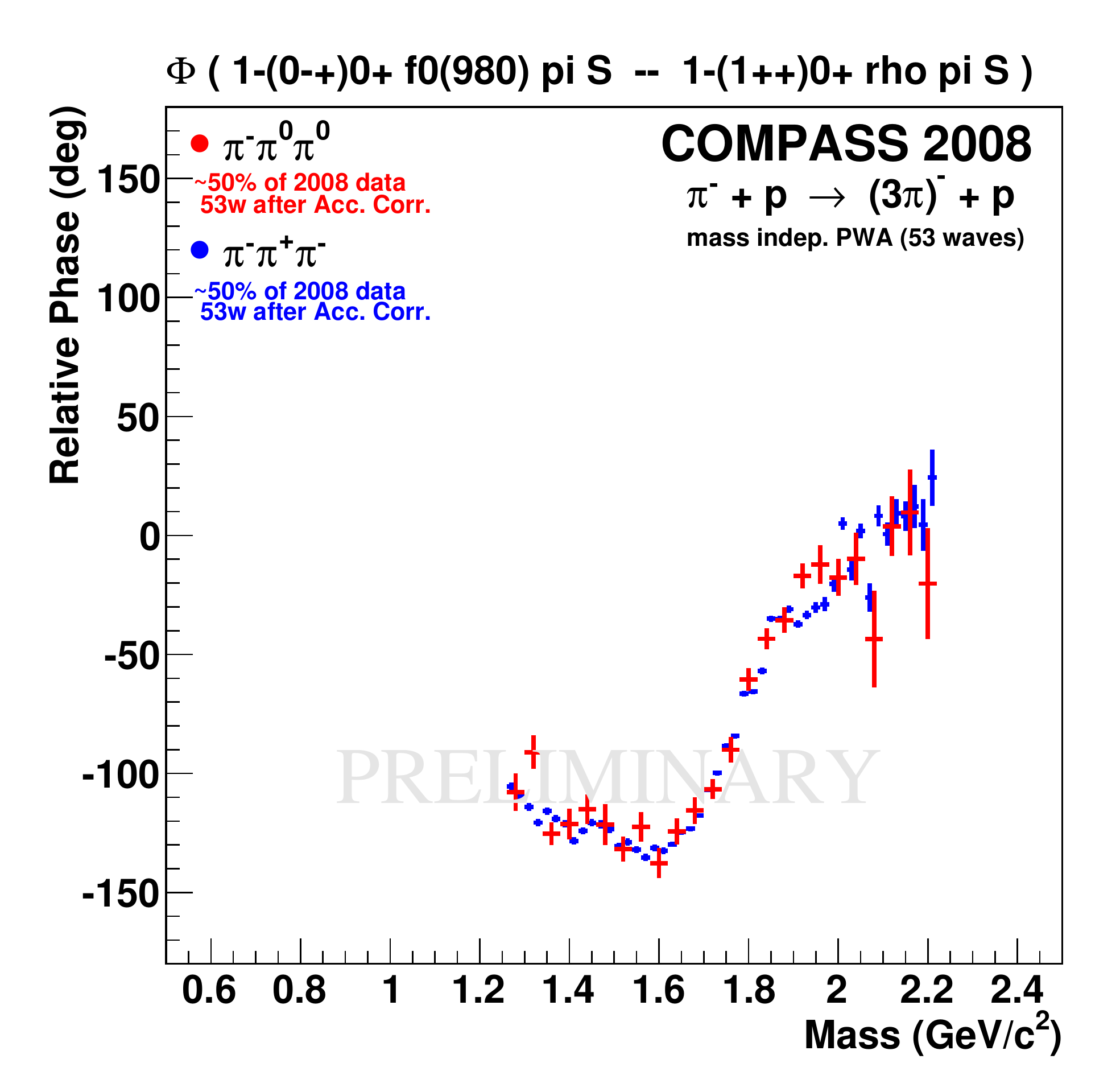} }
    \end{center}
  \end{minipage}
  \begin{minipage}[h]{.32\textwidth}
    \begin{center}
      \vspace{-0.5cm}
\resizebox{1.0\columnwidth}{!}{%
     \includegraphics[clip,trim= 5 0 10 15, width=1.0\linewidth, angle=0]{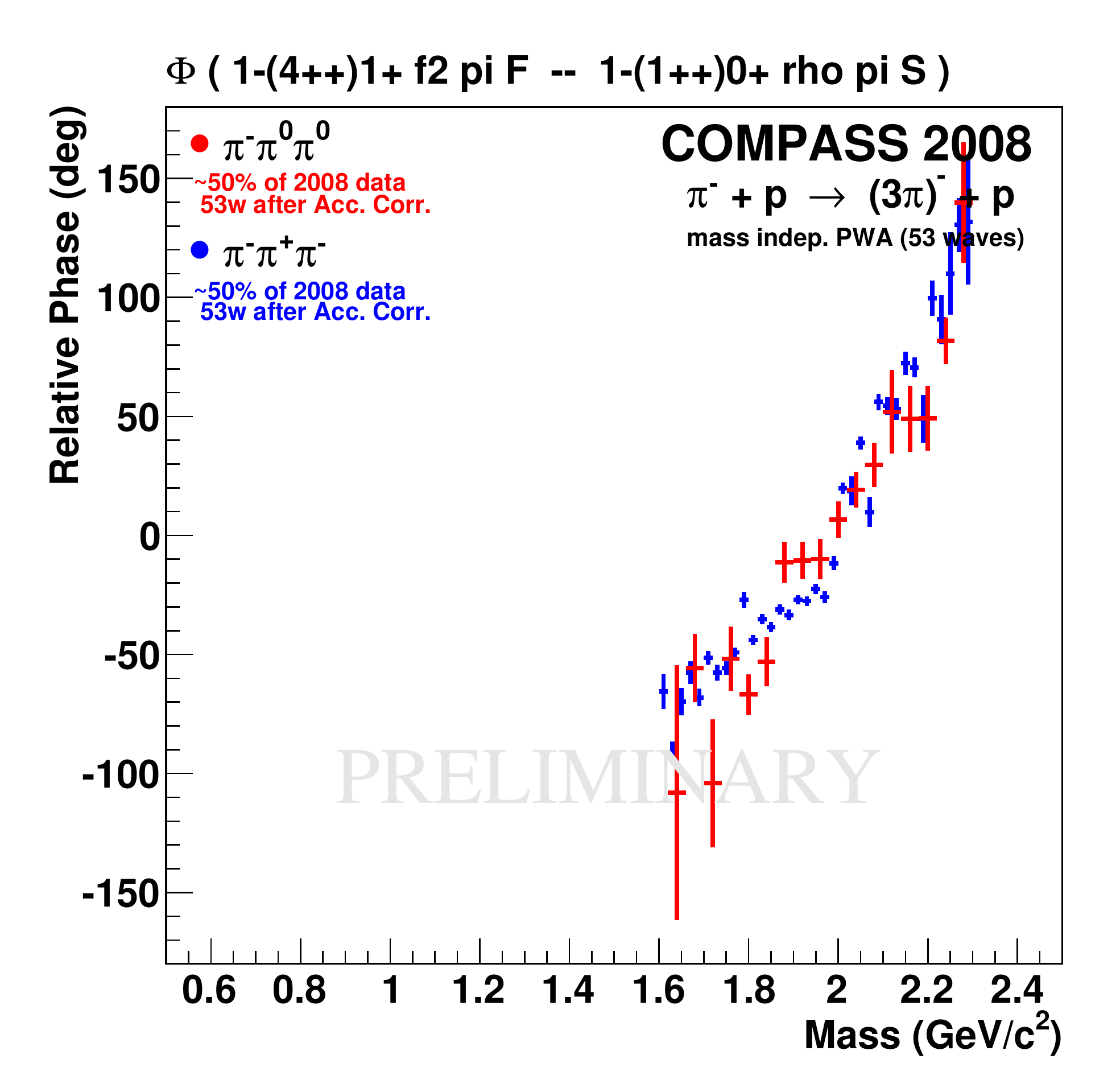}}
    \end{center}
  \end{minipage}
    \begin{center}
     \vspace{-0.3cm}
     \caption{Mass-independent PWA result for neutral (red) versus charged (blue) mode 
       --- small intensity waves and relative phases. The $a_1(1260)$ in the $(1^{++})0^+\,(\pi\pi)_s\pi\,P$ wave 
       {\it (top, left)}, the $\pi(1800)$ {\it (top, centre)} and the $a_4(2040)$ {\it (top, right)} are shown, 
       and the phase differences with respect to the $(1^{++})0^+ \rho\pi$ $S$ wave are given below, respectively.}
       \label{fig:phases}
     \end{center}
     \vspace{-0.7cm}
\end{figure}

More examples are given in Fig.\,\ref{fig:phases}, where the phase difference for a given wave with respect to the 
$a_1(1260)$ in the $(1^{++})0^{+}\,\rho\pi\,S$ wave (Fig.\,\ref{fig:isospinSymmMainWaves}, bottom/left) are given in addition. 
In the first example (Fig.\,\ref{fig:phases}, left), the phase difference is observed flat --- as expected, since
the two objects involved are the same and phase-locked due to the similar mass and width. For the other two, as the $\pi(1800)$
and the $a_4(2040)$ are rather separated in mass from the $a_1(1260)$ used as ``reference'', they are merely resonating 
against the tail of the $a_1(1260)$, manifesting in a clean, rapid phase motion restricted to the mass range of the observed
resonances. The phase differences are a very powerful tool to validate observed objects to be of resonant nature.  
The resultant phase motions are consistently observed, coinciding for the neutral and charged mode results.  
%
%
%
%

The mass-independent PWA result for the exotic $(1^{-+})1^+\,\rho\pi\,P$ wave, the fitted intensity and the 
phase difference again against the $a_1(1260)$, is shown in Fig.\,\ref{fig:exotic}, where the $(4^{++})1^+\,\rho\pi\,G$ 
wave of similar intensity showing the $a_4(2040)$ is re-displayed for comparison.
The neutral mode PWA result is compared to the charged one (Fig.\,\ref{fig:exotic}, centre) and for two 
different wave-sets (Fig.\,\ref{fig:exotic}, right) fitted to the data, comprising 42 and 53 partial-waves, 
respectively. We observe first of all about the same intensity for the neutral and the charged mode data (also the 
exotic wave obeys isospin symmetry). Further, we find two features for both modes on top of a relatively large, presumably 
non-resonant background that has the same shape in both cases. A larger peak appears at about 1.3\,GeV/$c^2$ 
and about 1.1\,GeV/$c^{2}$, respectively, for neutral and charged mode, which are still subject of detailed 
systematic studies (leakage, Deck, thresholds). Secondly, there is a smaller object at about 1.6\,GeV/$c^2$ 
that is consistently observed in the neutral and the charged mode results, just in the mass region where previous 
experiments reported the spin-exotic $\pi_1(1600)$ resonance.    

If we apply the same measure as for other small objects that confirms them to be resonances, as {\it e.g.} 
for the $a_4(2040)$ (Fig.\,\ref{fig:exotic}, left), namely looking at the relative phase against well 
established resonances like the $a_1(1260)$ (Fig.\,\ref{fig:exotic}, centre/bottom), we observe a clean, rapid phase 
motion exactly in the mass range of about 1.4 -- 1.8\,GeV/$c^2$, where the object is found in the intensity plot 
(Fig.\,\ref{fig:exotic}, centre/top) --- both the (even though small) signal and the rapid phase motion 
are observed consistently coinciding for both, neutral and charged mode results. Backgrounds appear differently in the 
phase differences below 1.4\,GeV/$c^2$ as in the fitted intensities.    
 
The PWA results presented in this paper are obtained fitting a wave-set of a total of 53 
partial-waves that is an extension of the 42 wave-set used previously~\cite{Alekseev:2009a,nerling:2009} by 
11 additional waves~\cite{nerling:2011} to account for the higher statistics analysed from the 2008 data. All resonances observed
(Figs.\,\ref{fig:isospinSymmMainWaves},\,\ref{fig:phases}) are similarly observed for both wave-sets, the results 
are robust and do not change with the wave-set extension. This holds also for the exotic wave, for which the 
results are shown for the neutral mode data for both PWA results, using the wave-set of 42 partial-waves (black 
points in Fig.\,\ref{fig:exotic}, right) or the extended one comprising 53 waves (re-displayed again as red points 
in Fig.\,\ref{fig:exotic}, right). Both, the signal at about 1.6\,GeV/$c^2$ as well as as the phase motion are 
observed rather unaffected for both wave-sets. 

In summary, we consistently reproduce the results from the 2004 data not only in the charged but also in the neutral 
mode 2008 data. As further systematic studies are ongoing (backgrounds from Deck, leakage), we do not yet draw strong 
conclusions on the existence of the $\pi_1(1600)$. Moreover, the charged data of huge statistics allow for much 
deeper studies in terms of a two dimensional ansatz, fitting the data simultaneously in bins of the $(3\pi)^{-}$ mass 
and the momentum transfer $t'$ allows for deeper understanding of {\it e.g.} backgrounds from the Deck effect.     
\vspace{-0.1cm}

\end{document}